\documentclass[conference]{IEEEtran}
\usepackage{makecell}
\IEEEoverridecommandlockouts
\usepackage{cite}
\usepackage[colorlinks,
            linkcolor=blue,
            anchorcolor=blue,
            citecolor=blue]{hyperref}
\usepackage{amsmath,amssymb,amsfonts}
\usepackage{algorithmic}
\usepackage{graphicx}
\usepackage{float}
\usepackage{textcomp}
\usepackage{xcolor}
\def\BibTeX{{\rm B\kern-.05em{\sc i\kern-.025em b}\kern-.08em
    T\kern-.1667em\lower.7ex\hbox{E}\kern-.125emX}}
\begin{document}

\title{RVRAE — A Dynamic Factor Model Based on Variational Recurrent Autoencoder for Stock Returns Prediction\\
}

\author{\IEEEauthorblockN{Yilun Wang}
\IEEEauthorblockA{\textit{Department of Economics} \\
\textit{North Carolina State University}\\
Raleigh, NC\\
ywang323@ncsu.edu}
\and
\IEEEauthorblockN{Shengjie Guo}
\IEEEauthorblockA{\textit{Department of Electrical and Computer Engineering} \\
\textit{North Carolina State University}\\
Raleigh, NC\\
sguo25@ncsu.edu}
}

\maketitle

\begin{abstract}
In recent years, the dynamic factor model has emerged as a dominant tool in economics and finance, particularly for investment strategies. This model offers improved handling of complex, nonlinear, and noisy market conditions compared to traditional static factor models. The advancement of machine learning, especially in dealing with nonlinear data, has further enhanced asset pricing methodologies. This paper introduces a groundbreaking dynamic factor model named RVRAE. This model is a probabilistic approach that addresses the temporal dependencies and noise in market data. RVRAE ingeniously combines the principles of dynamic factor modeling with the variational recurrent autoencoder (VRAE) from deep learning. A key feature of RVRAE is its use of a prior-posterior learning method. This method fine-tunes the model's learning process by seeking an optimal posterior factor model informed by future data. Notably, RVRAE is adept at risk modeling in volatile stock markets, estimating variances from latent space distributions while also predicting returns. Our empirical tests with real stock market data underscore RVRAE's superior performance compared to various established baseline methods.
\end{abstract}


\section{Introduction}
Forecasting financial market dynamics has become a central focus of modern market research, particularly in the challenging task of predicting stock market returns. Traditional asset pricing models, notably static factor models, encounter difficulties when dealing with high-dimensional datasets, and noisy financial data. In recent years, machine learning, especially deep learning, has emerged as a promising method to solve above problems due to its ability to handle vast amounts of data and complex nonlinear relationships. Recent studies \cite{b1},\cite{b2} highlight the potential of deep learning models in overcoming these limitations and providing more accurate return predictions.

Traditional static factor model, as proposed by \cite{b3}, $r_{t} = ~{\beta_{i}}^{'}f_{t} + u_{t}$, where  t = 1, …, T, $f_{t}$ is risk premia and  $\beta_{i}$ is factor exposures, and $u_{t}$ is the error term, establishes a linear relationship between stock returns and corresponding risk premia. However, challenges arise with dynamic betas, indicating that factor exposures may change over time as firms evolve. Latent factor models address this limitation by introducing dynamic betas. The form of a latent factor model is expressed as\begin{equation}
    r_{t} = ~{\beta\left( z_{t - 1} \right)}^{'}f_{t} + u_{t}
\end{equation}Where factor exposures are based on a high dimensional vector of asset characteristics $z_{t-1}$. 
The problem has transferred to how to identify the nonlinear relationship between bata and characteristics. Traditional statistical models cannot solve it. Recently, Deep learning offers a data-driven perspective to latent factor models, automating the extraction of latent factors from market data. ML solutions, such as those proposed by \cite{b4} and \cite{b5}, demonstrate effectiveness in automatically extracting latent factors, surpassing traditional methods in real-market scenarios.

However, most existing deep learning literature often overlooks temporal dependencies in time series data where the value at a current time point depends on previous values, and not pay enough attention to complex and noisy market which has a very low signal-to-noise ratio. Not considering intersections between time series data will introduce potential biases. For those who apply Recurrent Neural Network (RNN) and its variants (eg. GRU, LSTM), many scholars have shown its inability in working in a noisy environment. \cite{b6} Moreover, the low signal-to-noise ratio in stock data poses a challenge for existing ML solutions, affecting the effectiveness of latent factors extraction and hindering the development of an effective factor model for predicting cross-sectional returns. We propose RVRAE, an innovative latent dynamic factor model, which combine Recurrent Neural Network (RNN) and variational autoencoder (VAE). The former has proven its strong ability to solve temporal dependency between time series data and the latter shows its advantage in factor extraction from noisy data. 

Our approach utilizes the variational recurrent autoencoder (VRAE) to handle factors as latent variables, modeling data noise through their distribution in VRAE's latent space. We employ a prior-posterior learning method for the effective extraction of factors crucial for predicting cross-sectional returns. The method leverages an encoder-decoder structure, which uses future stock returns to identify optimal factors for return reconstruction. A predictor, trained on historical data, then approximates these factors. In the prediction phase, only this predictor and decoder are active, preventing future data leakage. This model uniquely combines return prediction with probabilistic risk estimation, factoring in inherent randomness in stock returns.
The contributions of our paper are as follows:\begin{itemize}
    \item We introduce RVRAE as an innovative dynamic latent factor model tailored to extract key factors from noisy market data efficiently. This model integrates Recurrent Neural Networks (RNN) and Long Short-Term Memory (LSTM) architectures to capture the temporal dependencies in time series data. Additionally, it incorporates a Variational Autoencoder (VAE) to address challenges in markets with low signal-to-noise ratios.
    \item To the best of our knowledge, we are the first to apply VRAE to blend RNN and VAE in an encoder-decoder framework specifically for factor extraction in noisy markets, creating a probabilistic approach for both return and risk estimation.
    \item Our comprehensive experiments on actual stock market data reveal that RVRAE outperforms traditional linear factor models as well as various machine learning and deep learning-based models in predicting cross-sectional returns.
\end{itemize}

\section{Related Work}

\subsection{Dynamic Factor model}

Factor models, which aim to explain the variation in stock returns, can be broadly classified into two categories: static models and dynamic models. In static factor models, the factor exposure of a stock remains time-invariant. The foundational static factor model is the Capital Asset Pricing Model, proposed by \cite{b7}. However, traditional static factor model has been challenged by time-varying beta and noisy market data. \cite{b8}

In dynamic factor models, the factor exposure of stocks varies over time. This variation is often calculated based on firm or asset characteristics such as market capitalization, book-to-market ratio, and asset liquidity. Reference \cite{b4}  introduces instrumented principal components analysis into dynamic factor models, allowing factor exposure to depend partially on observable asset characteristics with a linear relationship.

Going beyond linear relationships, \cite{b9} proposes a latent dynamic factor asset pricing model incorporating a conditional autoencoder network. This model introduces non-linearity into return dynamics, addressing limitations found in linear factor models. The authors demonstrated that their non-linear factor model outperformed leading linear methods in terms of effectiveness.

\subsection{Recurrent Neural Network (RNN)}

However, challenges faced by factor models are focused on two sides: the presence of temporal dependency between stock return data and noisy market data with low signal-to-noise ratios. Traditional feedforward neural network (FFN) structure treats each input as independent and identical distribution, which neglects the intersection between time series data. Not considering temporal dependencies can lead to poor predictions and misunderstanding of the data. Therefore, there is ongoing research dedicated to addressing the issue of learning temporal dependency between time series data. 

In recent years, more and more scholars use Recurrent Neural Network (RNN) to explore temporal dependency in market data. Reference \cite{b10} explored the use of LSTM networks in predicting stock market movements and found that these models significantly outperform traditional time series models. \cite{b11} integrates sentiment analysis from financial news with RNNs to predict stock price movements. \cite{b12} focuses on the application of RNNs in high-frequency trading. They found that RNNs, particularly LSTMs, are highly effective in capturing the dynamics of the market at higher frequencies

\subsection{Variational Autoencoder (VAE)}

For non-sequential data, VAEs have recently been shown to be an effective modelling paradigm to recover complex multimodal distributions over the data space \cite{b13}. A VAE introduces a set of latent random variables z, designed to capture the variations in the observed variables x. 
The joint distribution is defined as:\begin{equation}
    p\left( {x,z} \right) = p\left( x \middle| z \right) p(z)
\end{equation}The prior over the latent random variables, $p(z)$, is generally chosen to be a simple Gaussian distribution and the conditional $p\left( {x,z} \right)$ is an arbitrary observation model whose parameters are computed by a parametric function of $z$. Importantly, the VAE typically parameterizes $p\left( {x,z} \right)$ with a highly flexible function approximator such as a neural network. While latent random variable models of the form are not uncommon, endowing the conditional $p\left( {x,z} \right)$ as a potentially highly non-linear mapping from $z$ to $x$ is a rather unique feature of the VAE.
However, introducing a highly non-linear mapping from $z$ to $x$ results in intractable inference of the posterior $p\left( {x,z} \right)$. Instead, the VAE uses a variational approximation $q\left( {z,x} \right)$ to approximate $p\left( {x,z} \right)$.

The approximate posterior is a Gaussian $N\left( {\mu,diag\left( \sigma^{2} \right)} \right)$ with the mean $\mu$ and variance $\sigma^{2}$ are the output of a highly non-linear function of $x$, once again typically a neural network. The generative model $p\left( {x,z} \right)$ and inference model $q\left( {z,x} \right)$ are then trained jointly by maximizing the evidence lower bound (ELBO) with respect to their parameters, where the integral with respect to $q\left( {z,x} \right)$ is approximated stochastically. The gradient of this estimate can have a low variance estimate by reparameterizing $z$.\begin{equation}
    E_{q{({z|x})}}\left( {logp\left( x \middle| z \right)} \right) = E_{p(\varepsilon)}\left( logp\left( x \middle| {z = \mu + \sigma\varepsilon} \right) \right)~
\end{equation} 

\subsection{Variational recurrent Autoencoder (VRAE)}

Though VAE has already shown its advantage in efficient learning of complex distribution, it is not inherently designed to handle sequential data, such as time series, music, or text. \cite{b14} and \cite{b15} propose that a new recurrent structure which contains a VAE at each time step and VAE is conditioned at the state variable $h_{t-1}$ of an RNN. The prior on the latent random variable is no longer a standard Gaussian distribution: \begin{equation}
    \left. Z \right.\sim N\left( \mu_{0,t},~diag\left( \sigma_{0,t}^{2} \right) \right)
\end{equation} where $\left\lbrack {\mu_{0,t},~\sigma_{0,t}} \right\rbrack = ~\varphi_{\tau}^{prior}\left( h_{t - 1} \right)$
Moreover, the generating distribution will not only be conditioned on $z$ but also on $h_{t-1}$.\begin{equation}
    \left. x_{t} \middle| z_{t} \right.\sim N\left( \mu_{x,t},~diag\left( \sigma_{x,t}^{2} \right) \right.
\end{equation} where $\left\lbrack {\mu_{x,t},~\sigma_{x,t}} \right\rbrack = ~\varphi_{\tau}^{dec}\left( \varphi_{\tau}^{z}{\left( z_{t} \right),~h}_{t - 1} \right)$
Then, RNN are used to update hidden states at the last step.\begin{equation}
    h_{t} = ~f_{\theta}\left( \varphi_{\tau}^{x}\left( x_{t} \right),~\varphi_{\tau}^{z}\left( z_{t} \right),h_{t - 1} \right)
\end{equation}

VRAE has shown its success in many industries. In recent years, VRAE's scope has expanded to include anomaly detection, with notable advancements in identifying irregular events in video data \cite{b16}, in monitoring anomalies in solar energy systems \cite{b17}, and in estimating stock volatility \cite{b18}.

\section{Methodology}
According to \cite{b19},\begin{equation}
    r_{t} = ~{\beta_{t}}^{'}f_{t} + u_{t}~~~~~
\end{equation} where where $t$ = $1, …, T$, $f_{t}$ is risk premia, $\beta_{t}$ is factor exposures, and $u_{t}$ is the idiosyncratic error. We use VRAE, which contains VAE in each timestep. The formulation of our task is to learn a dynamic factor model with parameter $\theta$ to predict future cross-sectional returns from historical data. 

Our network model is composed of two modules:beta network and factor network.

\begin{figure*}
\centering
\includegraphics[width=16cm]{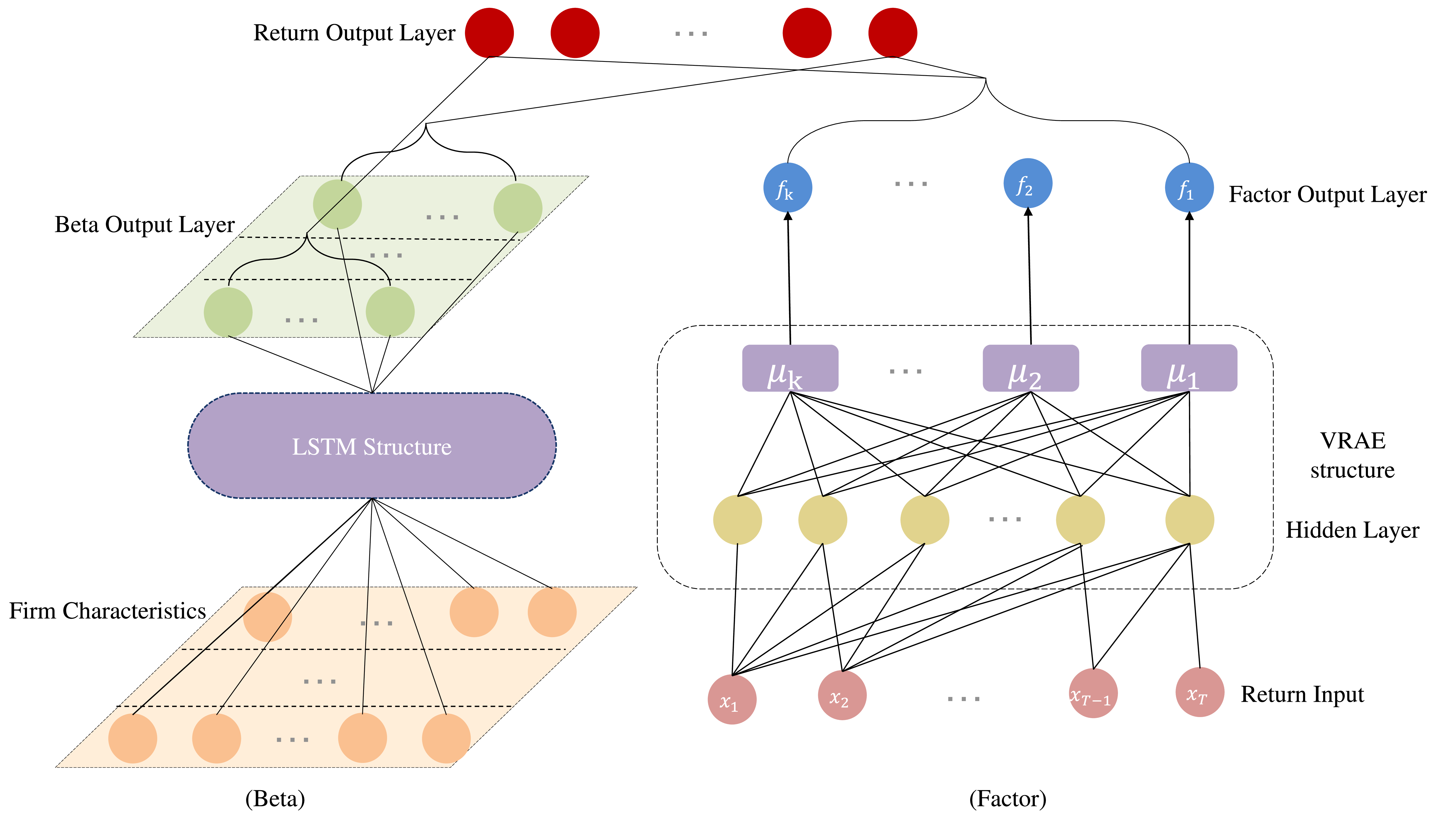}
\caption{RVRAE Model Structure}
\end{figure*}
In the first module about factor network, we combine  recurrent structure with variational autoencoder, assume that the vector of cross-sectional returns $r_{t}$ is driven by a vector of latent variables $z_{t}$ and hidden state variables $h_{t-1}$, then apply the regularized variational recurrent autoencoder (RVRAE) to learn the latent factor $f_{t}$ from the approximated posterior distribution of $z_{t}$. Specifically, $r_{t}$ as the input has the following marginal likelihood: \begin{equation}
    p_{\theta}\left( r_{t} \right) = {\int{p_{\theta}\left( r_{t} \middle| z_{t} \right) p_{\theta}\left( z_{t} \right)dz_{t}}}
\end{equation} where $p_{\theta}\left( z_{t} \right)$ is the prior probability. In vanilla VAE, $p_{\theta}\left( z_{t} \right)$ is assumed as a standard Gaussian distribution. Then we can get posterior probability of $z_{t}$:\begin{equation}
    p_{\theta}\left( z_{t} \middle| r_{t} \right) = ~\frac{p_{\theta}\left( r_{t} \middle| z_{t} \right) p_{\theta}\left( z_{t} \right)}{p_{\theta}\left( r_{t} \right)}
\end{equation} where $p_{\theta}\left( r_{t} \middle| z_{t} \right)$ is the likelihood function and $\theta$ is a vector of unknown parameters. Up till now, the method to compute $p_{\theta}\left( z_{t} \middle| r_{t} \right)$ is intractable, especially when $p_{\theta}\left( r_{t} \middle| z_{t} \right)$ has a complex structure. To solve the problem, we use a new function $q_{\varnothing}\left( \left. z_{t} \middle| r \right._{t} \right)$ with unknown parameters $\phi$ to approximate the true posterior $p_{\theta}\left( z_{t} \middle| r_{t} \right)$. During this process, given $r_{t}$ and an initial hidden state variable $h_{0}$, it produces a distribution over the possible values of $z_{t}$ from which $r_{t}$ could have been reconstructed. Thus, we can construct:\begin{equation}
    q_{\varnothing}\left( z_{t} \middle| r_{t} \right) = N\left( \mu_{zt},~diag\left( \sigma_{zt}^{2} \right) \right)
\end{equation} where $\left\lbrack {\mu_{zt},~\sigma_{zt}}\right\rbrack = ~\varphi_{\tau}^{encoder}\left( \varphi_{\tau}^{r}{\left( r_{t} \right),~h}_{t - 1} \right)$, $\mu_{zt}$, $\sigma_{zt}$ are outputs of neural network with returns $r_{t}$ and state variables $h_{t-1}$. $\varphi_{\tau}^{encoder}$ can be any highly flexible function such as neural networks and $\varphi_{\tau}^{r}$ means function can extract features from $r_{t}$.

Therefore, our encoder is as fllows:\begin{equation}
    h_{0} = 0,~r_{0} = ~r_{t}\label{con:11}
\end{equation}\begin{equation}
    h_{t} = tanh\left( w_{encoder}^{T}h_{t - 1} + w_{in}^{T}r_{t} + b_{encoder} \right)\label{con:12}
\end{equation}\begin{equation}
    \mu_{zt} = w_{\mu}^{T}h_{t} + b_{\mu}\label{con:13}
\end{equation}\begin{equation}
    log\left( \sigma_{z}^{2} \right) = w_{\sigma}^{T}h_{t} + b_{\sigma}\label{con:14}
\end{equation}\begin{equation}
    \left. Z \right.\sim N\left( \mu_{zt},~diag\left( \sigma_{zt}^{2} \right) \right)\label{con:15}
\end{equation}Equation (\ref{con:11}) initializes the network with the vector of individual asset returns $r_{t}$ and the state variables $h_{t}$. Equation (\ref{con:12}) shows how RNN updates its hidden state with recurrence equation. Equation (\ref{con:13}) and (\ref{con:14}) calculate the mean and the standard deviation of posterior factors from the mapping layer,respectively. Equation (\ref{con:15}) shows the distribution of $z_{t}$. 
Before we come to the decoder process, we move our eyes to the reparameterization trick. If we sample directly samples $z_{t}$ from $q_{\varnothing}\left( z_{t} \middle| r_{t} \right)$, the sampling behavior is undifferentiable. Thus, we use the reparameterization trick proposed by \cite{b14} to solve it. However, our model samples multiple $z_{t}$ at different timesteps, and for $q_{\varnothing}\left( z_{t} \middle| r_{t} \right)$, the mean and covariance do not directly depend only on $z_{t-1}$, but also $h_{t}$. After using the reparameterization trick,$z_{t}$ can be sampled as follows:\begin{equation}
    z_{t~} = ~\mu_{z}\left( h_{t},r_{t} \right) + ~{diag\left( {\sigma\left( h_{t},r_{t} \right)}_{z}^{2} \right)}^{\frac{1}{2}}\varepsilon_{t}
\end{equation} where $\left. \varepsilon_{t} \right.\sim N(0,~I)$, we can sample $z_{t}$ by just sampling $\varepsilon_{t}$, where $\varepsilon_{t}$ is viewed as the stochastic input of the
model with a standard multivariate normal distribution depending not on any unknown parameters. Then, we pass $z_{T}$ from the last timestep to decoder. 
For decoder process, the model needs simultaneously specify a probabilistic decoder $p_{\theta}\left( r_{t} \middle| z_{t} \right)$, which produces a distribution over the possible values of corresponding to a given code $z_{t}$.\begin{equation}
    p_{\theta}\left( r_{t} \middle| z_{t} \right) = ~N\left( \mu_{rt},~diag\left( \sigma_{rt}^{2} \right) \right)
\end{equation} where $\left\lbrack {\mu_{rt},~\sigma_{rt}} \right\rbrack = ~\varphi_{\tau}^{decoder}\left( {\varphi_{\tau}^{z}{\left( z_{t} \right),~h}_{t - 1}} \right)$, $\mu_{rt}$, $\sigma_{rt}$ are outputs of neural network with returns $r_{t}$ and state variables $h_{t-1}$. $\varphi_{\tau}^{decoder}$ can be any highly flexible function such as neural networks and $\varphi_{\tau}^{r}$ means function can extract features from $z_{t}$. Thus, we construct the decoder part:\begin{equation}
    z_{0} = ~z_{T}\label{con:18}
\end{equation}\begin{equation}
    h_{0} = tanh\left( w_{z}^{T}z_{T~} + b_{z} \right)\label{con:19}
\end{equation}\begin{equation}
    h_{t} = tanh~\left( w_{decoder}^{T}h_{t - 1} + w_{r}^{T}r_{t} + b_{decoder} \right)\label{con:20}
\end{equation}\begin{equation}
    \hat{f_{t}} = sigmod\left( w_{out}^{T}h_{t} + b_{out} \right)\label{con:21}
\end{equation}Equation(\ref{con:18}) and (\ref{con:19}) initializes the network with state variable $h_{t}$ and latent variable $z_{t}$. Then similar to the encoding process, RNN updates its hidden state with recurrence (\ref{con:20}). Equation (\ref{con:21}) shows how we obtain the output layer $f_{t}$.

The second module of our model is to estimate  $\beta_{t}$ from firm characters $x_{t}$ with LSTM structure. Compared with traditional Deep neural network, LSTM has advantage in identifying temporal dependency between time series data. We use LSTM structure to extract beta from firm characters. 
\begin{equation}
    h_{0} = 0,~C_{0} = 0
\end{equation}\begin{equation}
    i_{t} = \sigma\left( {w_{i}h_{t - 1} + U_{i}x_{t} + b_{i}} \right)
\end{equation}\begin{equation}
    f_{t} = \sigma\left( {w_{f}h_{t - 1} + U_{f}x_{t} + b_{f}} \right)
\end{equation}\begin{equation}
    o_{t} = \sigma\left( {w_{o}h_{t - 1} + U_{o}x_{t} + b_{o}} \right)
\end{equation}\begin{equation}
    g_{t} = tanh\left( W_{g}h_{t - 1} + U_{g}x_{t} + b_{g} \right)
\end{equation}\begin{equation}
    c_{t} = f_{t}\odot c_{t - 1} + i_{t}\odot g_{t}
\end{equation}\begin{equation}
    h_{t} = o_{t}\odot tanh\left( c_{t} \right)
\end{equation}\begin{equation}
    \hat{\beta_{t}} = h_{T}\label{con:29}
\end{equation} Through (\ref{con:29}), we obtain the factor exposure $\hat{\beta_{t}}$. By combining two modules together, we can obtain the predicted returns $\hat{r_{t}}$, where $\hat{r_{t}}$ = $\hat{\beta_{t}}$ $\hat{f_{t}}$. Fig.1 shows our model structure. The beta network (left panel) describes how different factor loadings $\beta$(in green) depend
on firm characteristics $x_{t}$ (in orange). The factor network (right panel) describes how the latent factors $f_{t}$ are obtained from an vector of asset returns $r_{t}$ (in lower red) via VRAE. The upper red nodes in output layer are computed by multiplying each row from the beta network with the vector of latent factors (in blue) from the factor network

\section{Estimation}

Our estimation consists of two parts. The first part is to train an optimal factor model, which is to reduce the reconstruction error, and the second part is to enforce the prior factors to approximate to the posterior factors with Kullback–Leibler (KL) divergence. Thus, the objective function of our model is: \begin{equation}
\begin{split}
    L\left( {\theta,\phi;r_{t}} \right) & = ~\frac{1}{L}{\sum\limits_{l = 1}^{L}\left\| {r_{t} - \hat{r_{t}}(\theta)} \right\|_{2}^{2}} \\ 
    & + \lambda KL\left( {q_{\varnothing}\left( \left. z_{t} \middle| r \right._{t} \right) \parallel p_{\theta}\left( \left. z_{t} \middle| r \right._{t} \right)} \right)   
\end{split}   
\end{equation}Where $\theta$ and $\phi$ include all unknown weights and bias parameters. $L$ is the number of Monte Carlo method ,and $\lambda$ is the weight of KL divergence loss.  
According to \cite{b20}, we can replace KL divergence loss with ELBO. Formally, given input $r_{t}$, the ELBO of our model is defined as:\begin{equation}
\begin{split}
   ELBO & = ~E_{z \sim q_{\varnothing}{({z_{t}|r}_{t})}}\left( {logp_{\theta}\left( r_{i,t} \middle| z_{T} \right)} \right) \\ &- ~\frac{1}{T}{\sum\limits_{t = 1}^{T}{KL\left( {q_{\varnothing}\left( \left. z_{t} \middle| r \right._{t} \right) \parallel p_{\theta}\left( z_{t} \right)} \right)}} 
\end{split}   
\end{equation}
The model passes a single $z_{t}$ at the final timestep of encoder to the decoder. Moreover, compared with vanilla function to calculate ELBO, there is a crucial difference that while existing models only impose KL regularization on the last timestep, we impose timestep KL regularization and average the KL loss over all timesteps, which allows more robust model learning and can effectively mitigate posterior collapse. Through the above objective function, we can use Adam algorithm to obtain unbiased $\theta$ and $\phi$.

\section{Empirical Result}

\subsection{Dataset }\label{AA}
We apply our model to analyze the US equity market. We collect monthly individual stock returns from CRSP The returns dataset starts from January 2000 to December 2020. The number of stocks is 5682. According to \cite{b20} and \cite{b21}, we choose 46 firm characteristics. Finally, we split data into 15-year training set, 3-year validation set and 3-year test set. We also take 5 as the number of factor to construct our model by following . 

\subsection{Model comparison}
We compare our model with other popular machine learning based dynamic factor models:
\begin{itemize}
\item Instrumented Principal Component Analysis (IPCA): a linear factor model with Intrumental variables, which is from \cite{b4}.
\item CA: Conditional Autoencoder model with latent factor model from \cite{b9}.
\item FactorVAE: Probabilistic Dynamic Factor Model Based on Variational Autoencoder from \cite{b22}.
\item ALSTM: LSTM with an attention layer for stock returns prediction from \cite{b23}.
\item Trans: A factor model based on Transformer architecture to predict stock returns from \cite{b24}. 
\end{itemize}

\subsection{Statistical performance}
Following \cite{b9} and \cite{b4}, the out-of-sample performance is evaluated by the total and predictive $R^{2~}$ below on the testing sample. The total $R^{2~}$ evaluates the model’s ability to characterize the individual stock riskiness and the predictive $R^{2~}$ measures model’s ability to explain panel variation in risk compensation.\begin{equation}
{R_{total}^{2} = 1 - \frac{\sum\left( r_{t} - {\hat{\beta^{'}_{t}}}\hat{\left. f_{t} \right)} \right.^{2}}{\sum r_{t}^{2}}}
\end{equation}\begin{equation}
    R_{pred}^{2} = 1 - \frac{\sum\left( r_{t} - {\hat{\beta^{'}}}_{t}\hat{\left. \mu_{t} \right)} \right.^{2}}{\sum r_{t}^{2}}
\end{equation}Where $\mu_{t}$  is the prevailing sample average of $f_{t}$ up to month $t$. Table 1 reports the total and predictive $R^{2~}$, and a higher $R^{2~}$ indicates a better statistical performance.

\begin{table}[htbp]
\caption{total and predictive $R^{2~}$}
\begin{center}
\begin{tabular}{|c|c|c|}
\hline
\textbf{Models} & \textbf{\textit{Total $R^{2~}$}}& \textbf{\textit{Predictive $R^{2~}$}} \\
\hline
& & \\[-6pt]
IPCA & 18.32 & 0.36 \\
& &  \\[-6pt]
CA & 17.52 & 0.74 \\
& & \\[-6pt]
FactorVAE & 18.71 & 0.92 \\
& & \\[-6pt]
ALSTM & 20.33 & 0.91 \\
& & \\[-6pt]
Trans & 19.23 & 1.46 \\
& & \\[-6pt]
RVRAE & \textbf{20.06} & \textbf{1.57} \\
\hline
\end{tabular}
\label{tab1}
\end{center}
\end{table}
Through Table 1, we can get following results:
\begin{itemize}
\item RVRAE achieves the best results in both Total $R^{2~}$ and Predictive $R^{2~}$, followed by ALSTM and Trans. 
\item Models that consider temporal dependency (FactorVAE, ALSTM, Trans, RVRAE) performs better than traditional machine learning model based on Deep Neural Network (CA) and traditional linear factor model (IPCA).
\item Factor models with linear structure (IPCA and ALSTM) can perform well in total $R^{2~}$, but their performance in Predictive $R^{2~}$ drops quickly. 
\end{itemize}

\subsection{Financial Performance}

Sharpe Ratio is commonly used to gauge the performance of an investment by adjusting for its risk. The higher the Shape Ratio is, the greater the investment return relative to the amount of risk taken, and the better the investment. Thus, we use out-of-sample Sharpe Ratio to evaluate the financial performance. 
\begin{equation}
    Sharpe~Ratio = ~\frac{r_{t} - r_{f}}{standard~deviation\left( r_{t} \right)}
\end{equation}
Where $r_{f}$ is the risk-free rate and we use monthly Treasury bill rates from the Kenneth French Data Library as the risk-free rate. 
Transaction costs are always important in financial investment. Following \cite{b9}, we consider two situations: Sharpe Ratio without transaction costs and Sharpe Ratio with 30 bps transaction costs. Table 2 summarizes models’ financial performance in the view of Sharpe Ratio.

\begin{table}[tp]
\caption{out-of-sample Sharpe Ratio}
\begin{center}
\begin{tabular}{|c|c|c|}
\hline
\textbf{Models} & \textbf{\textit{\makecell{Sharpe ratios \\ without transaction costs}}}& \textbf{\textit{\makecell{Sharpe ratios \\with 30 bps transaction costs}}} \\
\hline
& & \\[-6pt]
IPCA & 0.86 & 0.64 \\
& &  \\[-6pt]
CA & 1.12 & 0.96 \\
& & \\[-6pt]
FactorVAE & 1.92 & 1.43 \\
& & \\[-6pt]
ALSTM & 2.15 & 2.01 \\
& & \\[-6pt]
Trans & 12.03 & 1.92 \\
& & \\[-6pt]
RVRAE & \textbf{2.26} & \textbf{2.03} \\
\hline
\end{tabular}
\label{tab2}
\end{center}
\end{table}
From Table 2, we find that:\begin{itemize}
    \item RVRAE still shows the best results in both out-of-sample Sharpe Ratio without or with 30 bps transaction costs, with 2.26 and 2.01, respectively. Models with Attention mechanism (ALSTM, Trans) show the second best results and models with linear structure shows the worst prediction results.
    \item It is interesting to notice that all models perform worse when adding transaction costs, but models with Attention mechanism (ALSTM, Trans) have the smallest difference between Sharpe ratios without and with 30 bps transaction. This indicates future work can be continued on Attention mechanism in easing transaction costs for investments.
\end{itemize}
\begin{table*}[!htbp]
\caption{Robustness of all methods}
\begin{center}
\begin{tabular}{|c|c|c|c|c|c|c|}
\hline
\textbf{Models} &\multicolumn{2}{|c|}{\textbf{m = 50}} &\multicolumn{2}{|c|}{\textbf{m = 100}} &\multicolumn{2}{|c|}{\textbf{m = 150}}\\
\hline
\textbf{Name} & \textbf{\textit{Rank IC}}& \textbf{\textit{Rank ICIR}}& \textbf{\textit{Rank IC}}& \textbf{\textit{Rank ICIR}} & \textbf{\textit{Rank IC}}& \textbf{\textit{Rank ICIR}}\\
\hline

IPCA & 0.015(0.005) & 0.126(0.082)&  0.016(0.005) &0.130(0.073) & 0.016(0.005) & 0.133(0.070)\\
CA & 0.042(0.006) & 0.226(0.052)&  0.045(0.006) &0.262(0.035) & 0.045(0.005) & 0.268(0.030)\\
FactorVAE & 0.042(0.008) & 0.236(0.020)&  0.041(0.006) &0.233(0.018) & 0.041(0.005) & 0.235(0.018)\\
ALSTM & 0.028(0.005) & 0.166(0.023)&  0.030(0.011) &0.22(0.033) & 0.031(0.003) & 0.240(0.030)\\
Trans & 0.035(0.004) & 0.202(0.037)&  0.036(0.005) &0.232(0.036) & 0.036(0.004) & 0.257(0.037)\\
RVRAE & 0.055(0.008) & 0.295(0.039)&  0.053(0.004) &0.361(0.040) & 0.053(0.007) & 0.363(0.052)\\

\hline
\end{tabular}
\label{tab3}
\end{center}
\end{table*}
\subsection{Robustness}
In this part, our study assesses how well models perform when certain stocks are missing from the training data. We randomly omit $m$ stocks from the training set, then use this modified dataset to train the models. Next, we forecast stock returns on a new test dataset, focusing particularly on the returns of the omitted $m$ stocks. We measure the models' effectiveness on these missing stocks by calculating the Rank IC and Rank ICIR on the test data. Table 3 shows results for various numbers of omitted stocks, based on 10 random seeds. It reveals that our RVRAE model outperforms other baseline methods for all values of 'm'. This indicates that RVRAE is particularly adept at handling stocks it has not previously encountered, making it well-suited for real-market scenarios, such as predicting returns for newly issued stocks.

\section{Conclusion}
In this paper, we show how to learn an effective dynamic factor model with Variational autoencoder and recurrent structure for predicting cross-sectional stock returns. Specifically, in view of the noisy market environment, we propose a dynamic factor model based on variational recurrent autoencoder (VRAE). By treating factors as the latent random variables in VRAE, the proposed model with inherent randomness can model the noisy data and estimate stock risk. We extract useful factor exposures from a lot of firm characteristics and obtain factor loadings relying on the estimation of the conditional distribution of the returns by VRAE. Our statistical performance with higher total and predictive $R^{2~}$ and financial performance with higher out-of-sample Sharpe Ratio demonstrate the effectiveness of our model. Overall, the RVRAE model serves as a new tool for asset pricing, and its superior capacity allows us to get informative features from rich datasets.

\end{document}